\documentclass[screen=true, sigconf]{acmart}
\AtBeginDocument{%
  }

\copyrightyear{2027}
\acmYear{2027}
\setcopyright{cc}
\setcctype{by}
\acmConference[ASPDAC '27]{32nd Asia and South Pacific Design Automation Conference}{January 25--28, 2027}{Tokyo, Japan}
\acmBooktitle{32nd Asia and South Pacific Design Automation Conference (ASPDAC '27), January 25--28, 2027, Tokyo, Japan}
\acmDOI{10.1145/3840404.3848214}
\acmISBN{979-8-4007-2923-2/2027/01}

\usepackage{xcolor} 
\usepackage{soul} 
\usepackage{subcaption}
\usepackage{multirow}

\begin{document}

%%
%% The "title" command has an optional parameter,
%% allowing the author to define a "short title" to be used in page headers.
\title[Naviq: An Open-Source SoC Simulator]{Naviq: An Open-Source System-on-Chip Simulator for Rapid Prototyping, Verification, and Design Space Exploration}

\author{Scott Smith}
\authornote{These authors contributed equally to this project.}
\affiliation{%
  \institution{UIUC}
  \state{Illinois}
  \country{USA}
  }
\email{scottcs2@illinois.edu}

\author{Lukas Zscherpel}
\authornotemark[1]
\affiliation{%
  \institution{UIUC}
  \state{Illinois}
  \country{USA}
  }
\email{lukasez2@illinois.edu}

\author{Alyssa Huang}
\authornotemark[1]
\affiliation{%
  \institution{UIUC}
  \state{Illinois}
  \country{USA}
  }
\email{azh4@illinois.edu}

\author{Marissa Lanz}
\affiliation{%
  \institution{UIUC}
  \state{Illinois}
  \country{USA}
  }
\email{mlanz2@illinois.edu}

\author{Deming Chen}
\affiliation{%
  \institution{UIUC}
  \state{Illinois}
  \country{USA}
  }
\email{dchen@illinois.edu}

%%
%% The abstract is a short summary of the work to be presented in the
%% article.
\begin{abstract}
Modern FPGA SoCs are no longer simply collections of accelerators on top of a programmable fabric. They now combine programmable logic with hardened processors, networks-on-chip (NoCs), memory, and components that can dominate end-to-end performance. Evaluating individual components is insufficient; system performance depends on the complex interplay of components spread across the SoC. However, current FPGA SoC simulation flows provide limited support for this style of evaluation. Vendor simulations capture these interactions but are prohibitively slow for iterative design-space exploration. At the same time, generic software models overly abstract the hardware components that determine real behavior, resulting in less accurate simulations. Fast, easy-to-use simulators that maintain strong accuracy to real SoCs are critical for enabling system debugging and design space exploration.

We present Naviq, an open-source, cycle-level software simulator for FPGA SoCs, tuned to AMD Versal-class architectures. Naviq bridges high-fidelity but slow vendor simulation and fast but overly abstract software modeling: it captures key behaviors of modern commercial FPGA SoC infrastructure while remaining fast enough for debugging and design-space exploration. Naviq models hardened SoC infrastructure in C++ and integrates user accelerators through Verilator, allowing designers to evaluate custom SystemVerilog in a full-system environment. At the center of Naviq is an AXI4-compliant NoC model that captures transaction ordering, packetization, routing, backpressure, and memory interactions. Naviq provides CPU and memory integration, protocol checking, checkpointing, tracing, and performance-analysis tools for debugging and design-space exploration. Experimental results show that Naviq closely reproduces the bandwidth and latency of Vivado’s RTL NoC model while accelerating the evaluated simulations by 81×–182×. We present five case studies demonstrating Naviq's ability to run full-system simulations, analyze the performance impact of routing and placement decisions, and help diagnose deadlocks and performance-related bugs. By combining calibrated platform fidelity, RTL integration, and open-source accessibility, Naviq enables practical full-system evaluation of modern FPGA SoC designs.
\end{abstract}

%%
%% The code below is generated by the tool at http://dl.acm.org/ccs.cfm.
%% Please copy and paste the code instead of the example below.
%%


%%
%% This command processes the author and affiliation and title
%% information and builds the first part of the formatted document.
\maketitle

\section{Introduction}
\label{sec:intro}

Modern FPGA SoCs combine programmable logic with hardened processors, memory controllers, and networks-on-chip (NoCs). These components introduce platform-level effects that can dominate performance in the deployed system. Accelerators that perform well in isolation may encounter bottlenecks when requests contend for shared interconnect and memory resources. Platform-aware simulation is therefore essential for both architectural exploration and pre-deployment debugging, but it must balance accuracy, runtime, and support for heterogeneous components.

Existing simulation flows do not fully satisfy these requirements. Vendor RTL simulation, such as that provided by the Vivado Design Suite~\cite{amd_vivado}, includes detailed models of FPGA IPs and hardened infrastructure~\cite{amd_pg453_axi_switch, amd_pg313_noc}, but is often too slow for iterative full-system evaluation. In our experiments, using Vivado to simulate the transfer of only 1~MB between just two endpoints over the NoC requires tens of minutes (see $\S$\ref{sec:motivation}). More realistic end-to-end experiments may require multiple GBs of simulated data movement, requiring much longer simulations.  This makes simulating large workloads and iterative design-space exploration impractical. 

Faster software simulators target adjacent problems. General-purpose architectural simulators such as gem5~\cite{gem5, lowepower2020gem5simulatorversion200} provide mature processor, cache, memory, and full-system software models, while other tools focus on GPUs, heterogeneous architectures, accelerator modeling, or trace-driven analysis~\cite{gem5-gpu, gpgpusim_1, gpgpusim_2, 10.1145/3472456.3472461, 7372595, PAAS, 6853196, 7783751}. However, these tools are not designed to reproduce the substrate of commercial FPGA SoCs. Modern SoCs use AXI4~\cite{AMBA, AMBA_AXIS} memory-mapped and streaming interfaces to connect RTL accelerators, processors, DMA engines, and memories through hardened NoCs. Accurate evaluation therefore requires modeling AXI4 transactions, endpoint behavior, packetization, routing, ordering, backpressure, and memory interactions while integrating the RTL that attaches to this infrastructure. Designers consequently face a tradeoff: vendor simulation captures these effects but is slow, while faster software models often omit them.

We present \textit{Naviq}, an open-source, cycle-level software simulator for FPGA SoCs, tuned to AMD Versal-class devices~\cite{amd_alveo_v80}. Naviq uses fast C++ models for hardened infrastructure while preserving Verilator-based integration of user RTL. Its central component is an AXI4 NoC model that captures transaction ordering, packetization, routing, flow control, backpressure, and memory-system interactions. Naviq supports connecting this model to gem5 processors and configured DDR/HBM models and provides automated RTL integration, AXI4 protocol checking, tracing, checkpointing, traffic generation, and performance-analysis tools. This hybrid approach enables full-system debugging and design-space exploration while running orders of magnitude faster than vendor RTL simulation.

This work includes the following contributions:

\textbf{Accurate full-system FPGA SoC modeling.}
Naviq spans programmable logic, hardened NoC infrastructure, processors, DRAM, and HBM. Across held-out randomized tests, Naviq achieves per-category average accuracies of at least 97.6\% for latency and 99.9\% for bandwidth relative to Vivado’s RTL NoC model.

\textbf{Fast hybrid C++/RTL simulation.}
Naviq models hardened infrastructure in C++ while integrating SystemVerilog accelerators through an automated flow that finds NoC-attached RTL components and generates their C++ shims. Naviq achieves category-average speedups of 81$\times$--182$\times$ over Vivado RTL simulation.

\textbf{System-level design-space exploration.}
Naviq supports configurable placement and routing, parameter sweeps, and performance analysis. Our HBM performance study demonstrates how poor NoC placement and routing decisions can result in a 7.0$\times$ reduction in aggregate throughput due to congestion and contention for limited link bandwidth. Two additional studies show that lower average hop count does not necessarily imply lower tail latency.

\textbf{Full-system debugging and verification.}
Naviq provides AXI4 protocol checking, tracing, checkpointing, performance diagnostics, and traffic visualization. Two case studies demonstrate how Naviq can help diagnose both deadlocks and performance-related bugs.
\section{Background}
\label{sec:background}

\subsection{NoCs and AXI4}

Modern FPGA SoCs integrate programmable logic with processors, DRAM/HBM controllers, and hardened NoCs. Unlike soft-logic interconnects \cite{amd_smartconnect}, which face resource and scalability challenges at high throughput~\cite{PANIC,SuperNIC}, hardened NoCs provide scalable connectivity without consuming programmable-logic resources~\cite{liu2025demystifyingfpgahardnoc}.  The Versal NoC~\cite{amd_pg313_noc} connects programmable logic, processors, and memory controllers through a configurable packet-switched network exposed through AXI4 Memory-Mapped (AXI4-MM) and AXI4-Stream (AXI4-S) interfaces. Accelerator performance depends on endpoint placement, routing, transaction ordering, backpressure, and memory contention. Simpler simulators may omit or simplify these interactions. Naviq models these details at the cycle level, resulting in highly accurate simulations. 

\subsection{System Simulators}

SoC simulation tools operate at different levels of abstraction. Vendor RTL simulation flows, such as Vivado~\cite{amd_vivado}, offer detailed models for FPGA IP blocks and hardened platform components, including the NoC. These flows are valuable for functional validation and for understanding how user designs interact with vendor infrastructure, but they can be prohibitively slow for full-system workloads (see $\S$\ref{sec:motivation}) and provide limited flexibility for large design-space studies.

Open-source RTL simulators such as Verilator~\cite{verilator} provide faster simulation for synthesizable SystemVerilog designs and are widely used in open-source projects~\cite{opentitan, asanovic2016rocket}. However, they cannot replace vendor flows when key platform components, such as hardened NoCs, are available only as encrypted models. Naviq instead combines Verilated designs with software models of SoC infrastructure.

Full-system architectural simulators provide another point in the design space. gem5~\cite{gem5, lowepower2020gem5simulatorversion200} provides mature models for processors, caches, memory systems, and system-level execution. Garnet~\cite{garnet, heterogarnet}, integrated with gem5, provides a detailed cycle-level NoC model with routers, virtual channels, and links. However, these tools are not designed to model FPGA SoC platforms directly: they do not natively support AXI4-MM and AXI4-S, FPGA NoC endpoints, or arbitrary RTL accelerator integration.

Several prior systems integrate accelerators or RTL models into full-system simulation. PAAS~\cite{PAAS} targets CPU-FPGA heterogeneous systems, while gem5+RTL~\cite{10.1145/3472456.3472461} integrates RTL models into gem5 through an RTL-object abstraction and model-specific wrappers. These systems are complementary to Naviq: they focus on accelerator/software interaction and RTL integration, but do not target calibrated reproduction of commercial FPGA SoC NoC behavior, including Versal-style AXI4 endpoints, routing, backpressure, DDR/HBM interactions, and vendor-matched latency/bandwidth.

The most closely related simulator to Naviq is RAD-Sim~\cite{RADSim}, which builds on BookSim~\cite{booksim} to support simulation of reconfigurable accelerator systems with AXI4-compliant NoCs. However, unlike Naviq, RAD-Sim is a generic NoC exploration framework rather than a simulator tuned to a commercial FPGA SoC platform: it does not integrate processor models, target calibrated Versal-class behavior, or support hierarchy-discovered NoC connections. 
\section{Motivation}
\label{sec:motivation}

\begin{figure}
\centering
\includegraphics[width=0.80\linewidth]{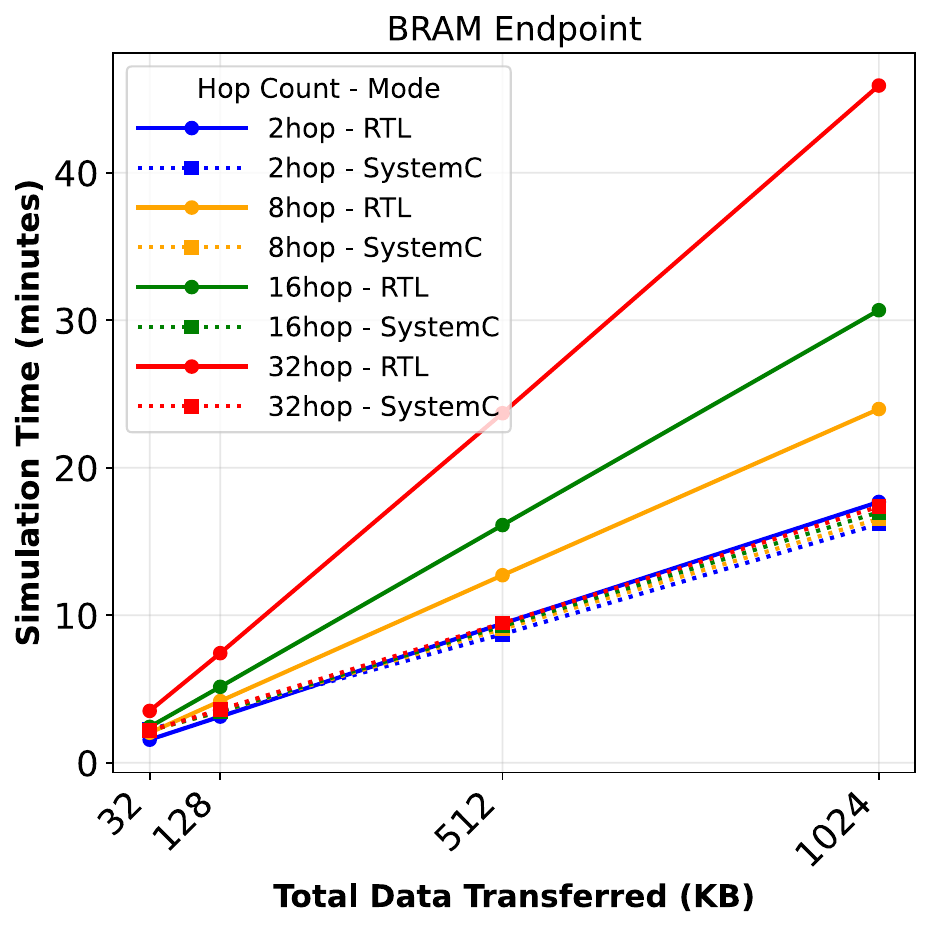}
\caption{Vivado Simulation Runtime vs Transferred Data}
\label{fig}
\end{figure}

Realistic FPGA SoC simulation requires practical runtimes, models for heterogeneous components (processors, memories, NoCs, and RTL), calibration against real platforms, and sufficient visibility for debugging. Existing tools partially satisfy these requirements. AMD's Vivado provides detailed models for platform components grounded in real hardware, but RTL simulation can be prohibitively slow and proprietary IPs limit visibility into internal behavior.

To quantify this cost, we ran Vivado 2025.1 on a server with a 24-core Intel Xeon 8168 processor and 64 GB of memory to simulate AXI4-MM traffic between a traffic generator and a BRAM endpoint. We varied hop count, transferred data, and the NoC model (Vivado RTL or SystemC). Figure~\ref{fig} demonstrates that runtime increases with both transferred data and hop count; in the 32-hop, 1 MB case, the RTL and SystemC models require approximately 45 and 15 minutes, respectively. This is a simple two-endpoint experiment; realistic systems with processors, memory, and accelerators will likely run even slower. Furthermore, Vivado's faster SystemC model is documented as providing less accurate performance numbers \cite{amd_pg313_noc} than the RTL model, which is within 5\% of the real system. The manual recommends using the RTL model when analyzing system performance. This presents system designers with an undesirable tradeoff between simulation accuracy and simulation speed. These limitations make Vivado poorly suited for design-space exploration.

Naviq  builds on gem5~\cite{gem5, lowepower2020gem5simulatorversion200} and Garnet~\cite{garnet, heterogarnet} to create a comprehensive solution. gem5 provides mature processor, cache, and memory models capable of executing realistic workloads, while Garnet provides a flexible cycle-level NoC foundation. Naviq adds the missing FPGA SoC capabilities: AXI4 transaction modeling, Versal-style NoC behavior, RTL integration, platform calibration, and debugging/design-space exploration tools. Together, these enable Naviq to accurately and quickly model current FPGA SoCs.
\section{The Naviq Simulator}
\label{sec:naviq_simulator}

\begin{figure}
    \centering
    \includegraphics[width=\linewidth]{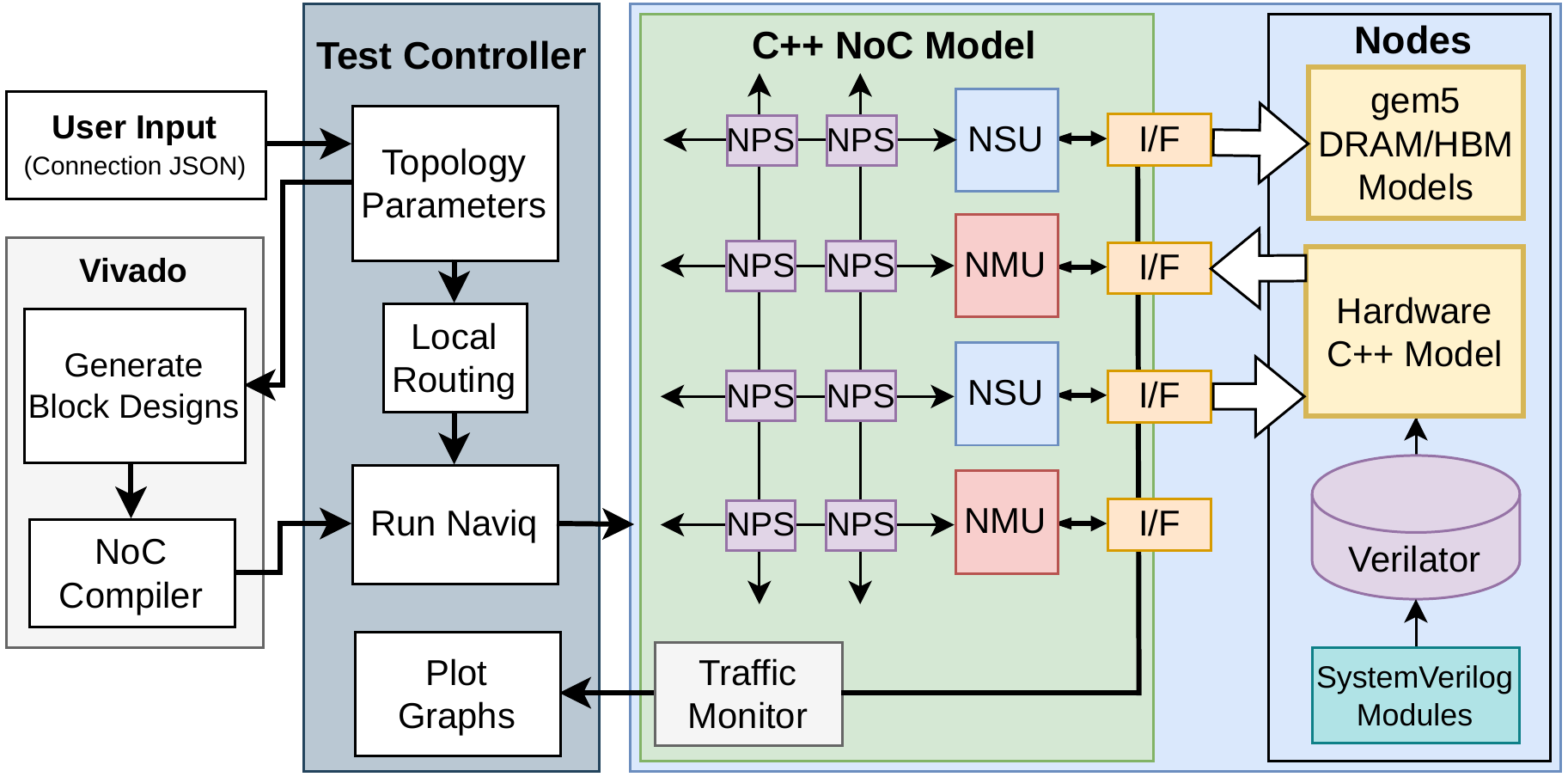}
    \caption{The Naviq Simulator and Associated Tooling}
    \label{fig:naviq_simulator}
\end{figure}

Naviq is a fast, cycle-level, hybrid C++/RTL simulator for modern FPGA SoCs. Naviq's hardware models are grounded in the Versal architecture \cite{amd_pg313_noc}. A high-level diagram is shown in Figure \ref{fig:naviq_simulator}. 

\subsection{The NoC Model}
Naviq overhauls Garnet's NoC \cite{garnet} to support AXI4.  Implementation and architectural choices described in this section reflect real Versal NoCs \cite{amd_pg313_noc}.  We adopt Versal terminology for the following NoC components: NoC Master Unit (NMU), NoC Slave Unit (NSU), and NoC Packet Switch (NPS).

\subsubsection{Nodes}
Nodes are any components attached to the NoC. These can range from gem5 memory models (HBM/DRAM) to custom accelerators written in SystemVerilog. Nodes can be attached to one or more NoC endpoints.

\subsubsection{NoC Endpoints}
\label{sec:design:subsec:noc_endpoints}

NoC endpoints are the ingress and egress points of the NoC. NMUs and NSUs enable AXI4 masters and slaves to connect to the NoC, respectively. Naviq supports a configurable number of endpoints, enabling experiments to scale from small tests to large NoC topologies. Each endpoint is configured for either AXI4-MM or AXI4-S communication. Endpoints translate AXI4 transactions to and from NoC flits, injecting and receiving at most one flit per cycle. For AXI4-MM, each NSU is assigned a unique address range, and NMUs route requests by translating addresses into NoC destinations; for AXI4-S, NMUs use the \textsc{TDEST} field to select the target NSU or fall back to a statically configured destination.

NMU write buffers limit the number of outstanding write transactions, and Naviq models this by tracking pending writes and gating ready signals when the buffer is full. Similarly, the read reorder buffer (RROB) tracks outstanding read requests, allows responses to complete out of order internally, and returns them to the node in order. When the RROB reaches capacity, the NMU backpressures the connected node, reproducing the same capacity-driven flow-control behavior observed on the platform. Naviq additionally exposes configurable endpoint timing parameters, such as per-beat write-data and read-response delays, which can be tuned to match measured or documented platform latencies.

\subsubsection{NoC Interface}

Naviq introduces a configurable NoC interface to mediate between each attached node and its corresponding NoC endpoint. Refactored from Garnet's \textsc{TileController}, the NoC interface provides a clock-driven boundary that can interact directly with RTL nodes while preserving cycle-level behavior. Each interface connects one node to one endpoint, is configured for a single protocol and role, and can model clock-domain crossing overheads between the node and NoC domains.

Each cycle, the node and NoC interface exchange data structures containing AXI4 signals. The interface decomposes outgoing node state into AXI4 channel values, enqueues them into the appropriate virtual-network queues, and reconstructs incoming NoC messages into the node-facing output structure. To keep this mechanism protocol-agnostic, each interface uses a \textsc{ProtocolHandler} object that implements the protocol-specific translation logic. Naviq provides handlers for AXI4-MM and AXI4-S; additional protocols can be supported by implementing new \textsc{ProtocolHandler} subclasses.

\subsubsection{NoC Fabric}
Naviq uses Garnet's NoC fabric to model flit-level routing and transport after endpoints translate AXI4 transactions into NoC flits. The fabric models router-like NPSes, virtual channels, buffering, arbitration, and credit-based flow control. We model Versal NPS behavior ~\cite{amd_pg313_noc}, including least-recently-used packet arbitration and routing policies. 

\subsubsection{Latency Calibration.}
The AXI4 standard \cite{AMBA_AXIS, AMBA} and Versal documentation \cite{amd_pg313_noc} define the NoC's behavior, but do not expose the cycle-level timing needed to match real hardware. To calibrate Naviq, we ran controlled Vivado experiments and inspected waveforms at the visible NMU, NSU, and NPS boundaries. We began with 1:1 traffic-generator-to-BRAM tests, varying transaction size, burst length, endpoint placement, and route length to separate fixed endpoint delays from route-dependent fabric delays. We compared Naviq and Vivado, using mismatches to refine packetization timing, flit emission, response-path behavior, and NPS arbitration. After these cases matched, we extended the same process to other traffic patterns to capture congestion effects and merge-point contention. 

\subsection{Memory and Processor Models}

Naviq supports the gem5 \cite{gem5} processor, HBM, and DRAM models. gem5 CPUs execute software and issue memory requests; Naviq models AXI4 packetization and NoC transport, and gem5 memory controllers model backend DRAM/HBM timing. Naviq supports \texttt{TimingSimple}, \texttt{AtomicSimple}, and \texttt{O3} CPUs. \texttt{TimingSimple} is the default for NoC studies because requests naturally stall on cache misses, bridge capacity, NoC backpressure, and memory delay. CPU traffic enters the NoC through a \texttt{CpuNocBridge}, which translates gem5 packets into AXI4-MM bursts or narrow MMIO transactions and propagates outstanding-transaction backpressure to the CPU.

Naviq provides fixed-latency BRAM endpoints as well as DDR and HBM endpoints from the gem5 \texttt{MemCtrl} and \texttt{HBMCtrl} models. A shared AXI4 slave frontend models NoC-visible request, response, and controller-entry timing before issuing backend memory requests to gem5. Address mappings are derived from the configured topology, allowing CPU, DMA, and RTL nodes to access DDR, HBM, or BRAM through the NoC. Naviq also supports functional memory access for binary loading and fast-forwarding before switching to timed CPU, NoC, and memory simulation for measurement.

\subsection{Naviq Tooling and Features}
\label{sec:tooling}

\subsubsection{Automated SystemVerilog Co-Simulation}
\begin{figure}
\centering
\includegraphics[width=0.75\linewidth]{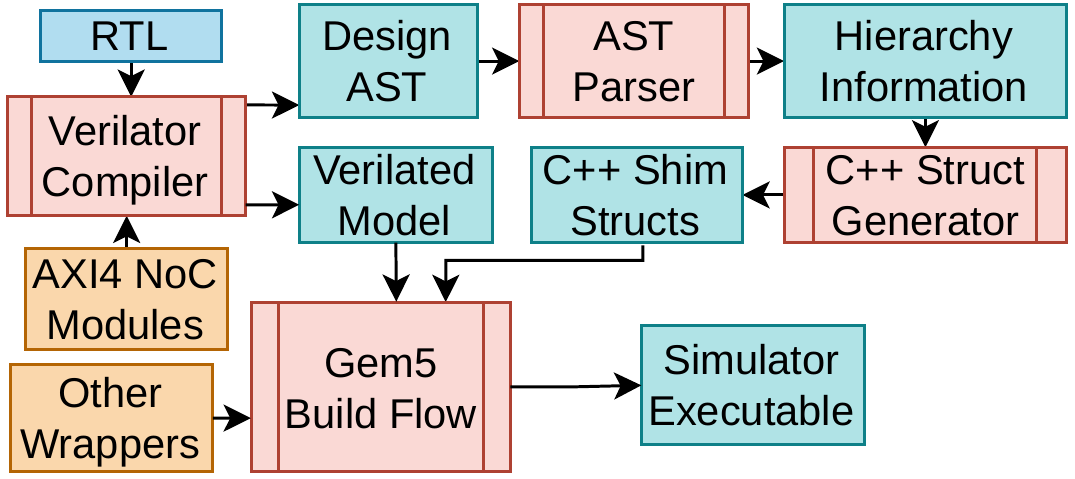}
\caption{Naviq's Automated RTL Integration Flow}
\label{fig:rtl_integration}
\end{figure}

SystemVerilog designs may implement equivalent functionality through different signal names, widths, and module hierarchies, making manually written simulation wrappers fragile to small code changes. To integrate custom RTL modules, Naviq requires only that the user RTL instantiate NMU and NSU modules, matching the
interface convention used by Vivado. As shown in Figure ~\ref{fig:rtl_integration}, Naviq invokes Verilator to generate both a C++ model and an abstract syntax tree (AST), uses the AST to locate NMU/NSU instances in the hierarchy, and automatically emits
C++ shim structures that access their signals. These shims are combined with Naviq's templated C++ libraries and compiled into the gem5-based simulator. This flow integrates existing RTL without hand-written C++ or SystemVerilog wrappers while supporting arbitrary module hierarchies.

\subsubsection{Configuration and Design-Space Exploration}
\label{sec}

To simplify experiment setup, Naviq provides a JSON-driven configuration flow. A connection configuration specifies the active endpoints, their parameters, and the logical connections among traffic generators, BRAM-like targets, CPU bridges, memory models, and custom RTL nodes. A separate placement configuration maps these endpoints onto physical NoC locations, allowing users to study placement effects without rewriting the logical system description. A Python launch script reads these configurations, constructs the corresponding gem5/Naviq system, and runs the selected workload. This design enables users to construct repeatable NoC experiments.

Internally, Naviq represents the NoC as a graph of physical resources, including packet switches, endpoint attachment points, and routing components. After placement, routes may be generated using built-in algorithms, user-provided routing logic, or imported Vivado-derived topology and route information. Naviq also provides reusable traffic generators, endpoint models, and test harnesses for controlled studies of traffic pattern, target bandwidth, burst structure, endpoint placement, memory attachment, and routing policy. A sweep controller expands parameterized experiments into multiple runs, aggregates selected performance metrics, and enables design-space studies in which one architectural variable is varied while the remaining system configuration is held fixed. Users may also bypass the JSON flow and construct custom systems directly through standard gem5-style Python configuration.

\subsubsection{Debugging and Protocol Verification}
\label{sec:graph}
To improve NoC visibility and debugging, Naviq provides built-in component tracing, packet snooping, and state checkpointing. Such mechanisms target two complementary classes of problems that may arise during full-system SoC simulations: protocol correctness violations that can be identified via explicit signal protocol requirements, and performance or design issues that must be inferred from system behavior. Naviq exposes multiple forms of system information, allowing users to investigate where and why unexpected behavior occurred.

Naviq's snooping infrastructure automatically monitors AXI4 traffic for protocol violations. AXI4 transactions are monitored as they traverse the NoC and are checked to ensure there are no violations of AXI4 properties that must hold. For example, Naviq can detect AXI4-MM ID mismatches, invalid ready/valid handshakes, and violations of AXI transaction-ordering requirements. These explicit correctness violation checks are particularly useful in full-system simulations, where a protocol error originating from one NoC endpoint may manifest as missing responses, stalled performance, or incorrect behavior. Rather than requiring the user to manually trace the transaction through the system to diagnose such bugs, Naviq points the user towards the first protocol-level event at which the behavior becomes invalid.

However, not all problematic behavior arises from protocol violations. A design may satisfy AXI4 requirements while exhibiting poor bandwidth utilization, excessive latency, congestion, or ineffective memory usage. To expose these issues,
Naviq traces activity from NoC components such as NPSes and NoC interfaces. These traces are used to generate bandwidth, latency, data-movement, and congestion statistics. Additionally, Naviq generates diagnostic warnings for certain conditions, such as unusually low HBM bandwidth or unexpectedly long latencies. These warnings help identify performance bugs. For example, low HBM bandwidth may result from unfair arbitration schemes or high HBM row-miss rates. These can be identified via Naviq's tracing features.

Debugging is costly when errors occur after millions of cycles. To remedy this, Naviq supports checkpointing. Naviq periodically records information about the system state, which can later be used to reconstruct the NoC at intermediate points of execution and continue execution. When an issue causes the simulation to hang or terminate with an error, a checkpoint near the onset of the bug can be used as a starting point for investigation, avoiding the need to reproduce the entire simulation trace from the beginning and consequently greatly reducing the time needed to localize a bug.

Together, these mechanisms provide multiple levels of visibility and debuggability into NoC and system-level behavior. Packet snooping addresses violations of explicit protocol requirements and can directly report correctness errors. Component tracing and graphing expose system behavior that helps diagnose performance issues. Checkpointing adds temporal localization by preserving intermediate states during simulation, decreasing the cost of investigating a bug during long-running simulations and allowing the user to split simulation execution into several sessions if needed. This combination allows Naviq to support both automated debugging and user-driven performance debugging, as well as providing full visibility into and control of simulation behavior that would otherwise be difficult to obtain during a full-system SoC simulation.
\section{Experiments}
\label{sec:experiments}
All experiments were run on a 24-core, Ubuntu 24.04 Intel Xeon 8168 server with
64 GB of memory and Vivado 2025.1.

\begin{figure}[t]
    \centering
    \includegraphics[width=\linewidth]{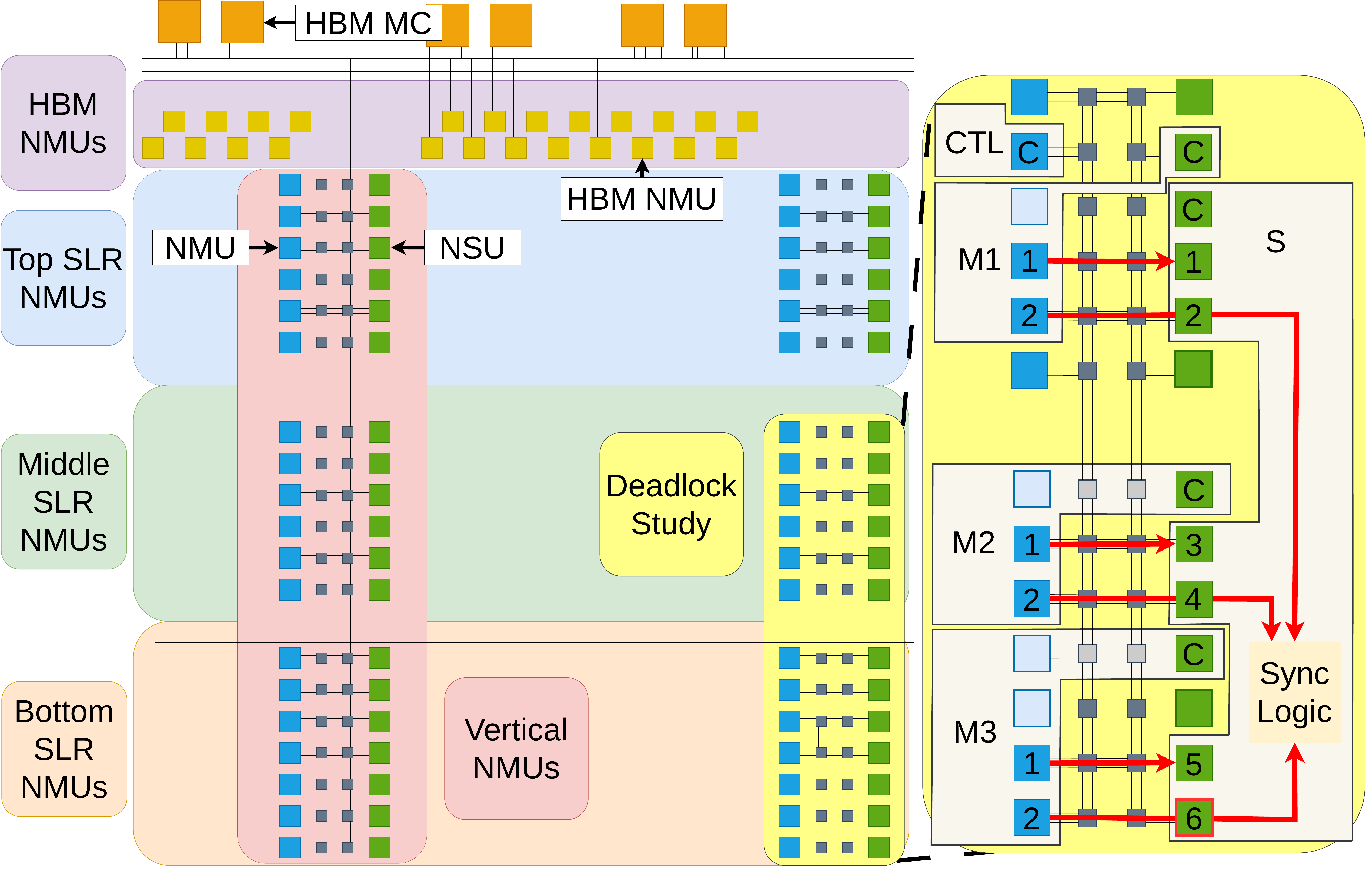}
    \caption{Logical View of the Naviq Case Studies}
    \label{fig:experiment-setup}
\end{figure}

\subsection{Naviq Accuracy and Speedup}

\begin{figure}[t]
    \centering
    \includegraphics[width=0.95\columnwidth]{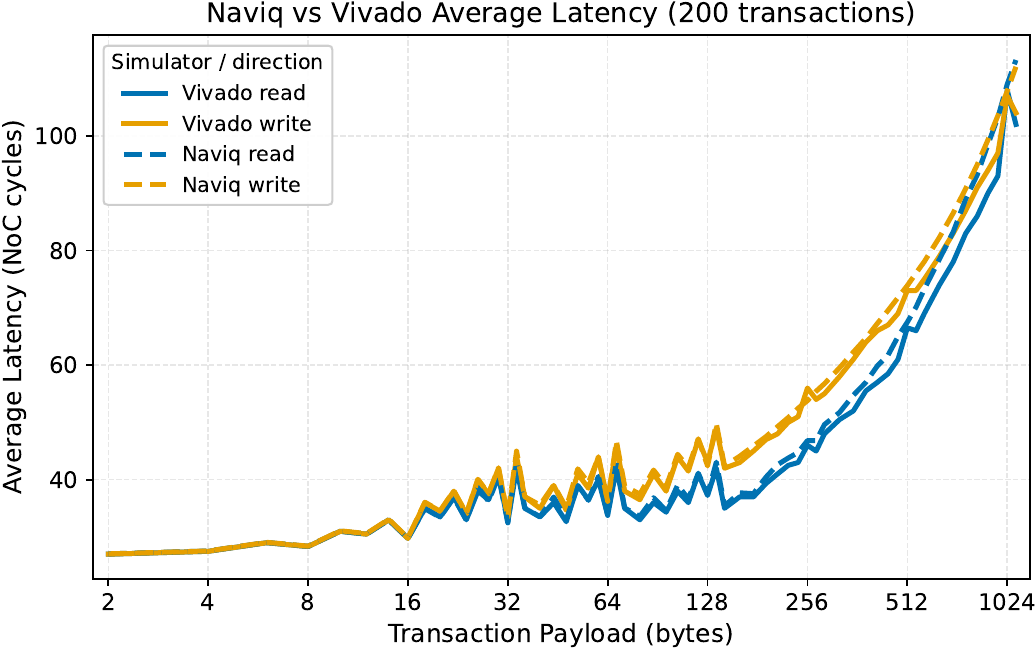}
    \caption{Naviq vs Vivado Average Transaction Latency}
    \label{fig:latency_plot}
\end{figure}

We evaluate Naviq’s accuracy against the Vivado RTL NoC model. First, we test point-to-point configurations using a fixed two-hop route and AXI4 transaction sizes ranging from 2~B to 1~KB. For each size, we issue 200 reads and 200 writes and measure the mean and maximum transaction latencies.  Figure~\ref{fig:latency_plot} plots the mean latencies. Averaged across sizes, Naviq’s mean read and write latencies differ from Vivado by only 1.03 and 1.04 cycles, respectively. Although not shown, the maximum latencies follow a similar trend, with average differences of 0.46 cycles for writes and 1.11 cycles for reads.

We then evaluate 20 more complex configurations: five 1:1, five 4:1 incast, five 1:4 outcast, and five 4:4 all-to-all cases. All use Vivado-generated topology and routing, 512-bit endpoints, 512~B transactions, 100 reads and 100 writes per generator, random endpoint locations, and a 1~GHz NoC clock. These tests were not used during calibration; all Naviq timing and architectural parameters were frozen before evaluation. Table~\ref{tab:naviq-category-averages} reports accuracy for the min, mean, and max observed latencies, together with bandwidth accuracy and simulation speedup. We define accuracy as
\begin{equation}
\mathrm{Accuracy}(x_{\mathrm{N}},x_{\mathrm{V}})
=
\left(
1 -
\frac{
\left|x_{\mathrm{N}}-x_{\mathrm{V}}\right|
}{
\max\left(x_{\mathrm{N}},x_{\mathrm{V}}\right)
}
\right)
\times 100\%,
\label{eq:accuracy}
\end{equation}
where $x_{\mathrm{N}}$ and $x_{\mathrm{V}}$ denote the corresponding
Naviq and Vivado measurements, respectively. Category-average accuracy remains at least 97.6\% for latency and 99.9\% for bandwidth, and category-average speedup ranges from 81$\times$ to 182$\times$ relative to Vivado.

Our accuracy evaluation compares Naviq's NoC latency and bandwidth against Vivado's RTL NoC model, while our speedup results compare Naviq against Vivado RTL simulation. The Vivado NoC documentation \cite{amd_pg313_noc} states that the Vivado RTL model is accurate to within 5\% of the real hardware. Naviq’s agreement with this model provides strong evidence that it captures realistic system behavior. The closest prior work, RAD-Sim~\cite{RADSim},
reports a 26$\times$ average speedup and 5.1\% average error for a
handwritten SystemC model of an NPU relative to RTL. It does not validate its BookSim-based \cite{booksim} NoC model against hardware or a commercial FPGA platform. In contrast, Naviq evaluates NoC latency and bandwidth against Vivado's RTL NoC model and achieves category-average speedups of 81$\times$--182$\times$. Because the two studies evaluate different components and use
different workloads and baselines, their numerical results are not necessarily directly comparable. We investigated an artifact-level comparison, but differences in traffic generation, RTL integration, and target-platform configuration made it difficult to perform a controlled apples-to-apples experiment.

\begin{table}
\centering
\caption{Naviq Bandwidth/Latency Accuracy and Speedup}
\label{tab:naviq-category-averages}
\small
\setlength{\tabcolsep}{3pt}
\begin{tabular}{ccccccc}
\toprule
\textbf{Category}
& \multirow{2}{*}{\shortstack{\textbf{Avg}\\\textbf{Hops}}}
& \multicolumn{3}{c}{\textbf{Latency}}
& \textbf{BW}
& \textbf{Speedup} \\
\cmidrule(lr){3-5}
& & \textbf{Min} & \textbf{Avg} & \textbf{Max} & & \\
\midrule
1:1 & 31.0 & 100.00\% & 100.00\% & 100.00\% & 100.00\% & 101.46$\times$ \\
1:4 & 18.1 & 100.00\% & 100.00\% & 99.57\% & 99.98\% & 182.02$\times$ \\
4:1 & 29.7 & 98.89\% & 97.67\% & 97.60\% & 99.99\% & 81.52$\times$ \\
4:4 & 20.4 & 97.88\% & 98.51\% & 98.70\% & 99.99\% & 123.64$\times$  \\
\bottomrule
\end{tabular}
\end{table}

\subsection{Full-System AES-to-HBM Placement Study}

\begin{figure}
    \centering
    \includegraphics[width=1\columnwidth]{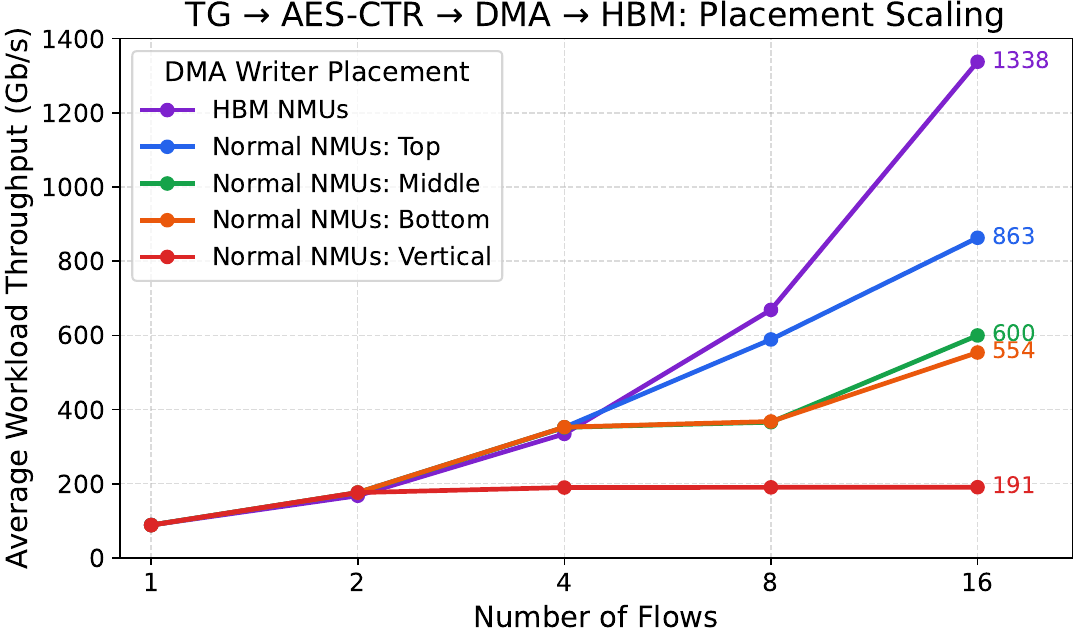}
    \caption{Aggregate HBM Throughput vs Flow Count}
    \label{fig:aes-throughput}
\end{figure}

We evaluate Naviq on a full-system encryption \cite{nist2001aes} experiment using a heavily modified AES encryption core from \cite{secworks_aes}. This modified core can sustain up to 200 Gb/s of AXI4-S traffic. The NoC endpoints sustain just over 100 Gb/s per NMU/NSU. This ensures the AES core is not a potential bottleneck; it can sustain whatever throughput the NoC provides. The full pipeline is Traffic Generators$\rightarrow$AES Cores$\rightarrow$DMA$\rightarrow$HBM. We evaluate 1--16 independent 1~MB flows. The AES core runs at 400~MHz, the NoC at 1~GHz, and the V80 HBM2e interface is configured to support just over 100 Gb/s per HBM NMU\footnote{Due to routing limitations, a maximum of 16 HBM NMUs were tested, but the simulator supports 32 per stack, matching the Versal NoC \cite{amd_pg313_noc} on the Alveo V80 \cite{amd_alveo_v80}.}. A simulated CPU configures per-flow AES/DMA state, programs descriptors and control registers, and launches the workload; after launch, each traffic generator feeds an AES-CTR RTL accelerator whose DMA writer sends encrypted data through an AXI4-MM NMU and the NoC to HBM. We measure aggregate write throughput from the release of the traffic generators to the final AXI4-MM write response, excluding CPU setup while preserving accelerator, DMA, NoC, route, and HBM behavior.

Figures~\ref{fig:experiment-setup} and~\ref{fig:aes-throughput} show that placement and routing strongly affect scaling. Non-vertical placements scale similarly through four flows, reaching 334--353~Gb/s, but diverge as NoC contention increases. At 16 flows, HBM-local NMUs reach 1.338~Tb/s, while top, middle, and bottom normal-NMU placements reach only 863, 600, and 554~Gb/s, respectively. The vertical placement, which forces all writers through one Vertical NoC column, saturates near 191~Gb/s from four flows onward. These results show that high-throughput FPGA SoC designs should exploit HBM-local NoC attachment when possible, place writers close to their target HBM resources, and avoid overlapping routes that concentrate traffic onto shared NoC columns or links. Overlapping routes increase congestion and competition for limited link bandwidth, hurting performance. This placement sensitivity is consistent with prior characterization of Versal NoCs~\cite{liu2025demystifyingfpgahardnoc}; Naviq exposes the same class of placement-sensitive effects during full-system simulation. This experiment demonstrates Naviq's ability to jointly simulate processor control software, Verilated accelerators, AXI4 NoC behavior, and HBM while exposing placement- and route-dependent contention.

\subsection{Route Overlap vs Hop Count Study}
\label{sec:route-overlap}

We evaluate Naviq's ability to compare NoC route configurations. Four AXI4-MM traffic generators communicate with BRAM-like targets using shift, reverse, tornado, and hotspot traffic patterns. For each pattern, we compare two route sets with lower and higher resource sharing, denoted low and high overlap, respectively. Endpoint placement, source-to-destination mapping, and traffic generator settings are identical across each test.

We evaluate configured read and write rates of 400 and
800~MB/s per source using matched random seeds across route
configurations. Achieved read and write bandwidth tracks the
configured rates, and all runs complete with the default
protocol-buffer capacity of 32 entries. For each endpoint and
direction, we pool transactions across seeds and compute P99
latency. Table~\ref{tab:route-overlap} reports the maximum of
these values across endpoints and both directions. The reported
values are identical at the two tested loads. This experiment
compares latency under controlled offered loads; it does not
measure maximum sustainable throughput.

\begin{table}[t]
  \centering
\caption{Average hops and worst-endpoint P99 latency (cycles)
for low- and high-overlap route sets.}
  \label{tab:route-overlap}
  \small
  \setlength{\tabcolsep}{4pt}
  \begin{tabular}{lrrrr}
    \toprule
    & \multicolumn{2}{c}{Average hops}
    & \multicolumn{2}{c}{Worst-endpoint P99} \\
    \cmidrule(lr){2-3}\cmidrule(lr){4-5}
    Pattern & Low & High & Low & High \\
    \midrule
    Shift   & 26.0 & 27.0 & 175 & 196 \\
    Reverse & 15.5 & 17.5 & 179 & 209 \\
    Tornado & 44.5 & 44.0 & 179 & 210 \\
    Hotspot & 37.0 & 36.0 & 302 & 298 \\
    \bottomrule
  \end{tabular}
\end{table}

The higher-overlap route set has higher worst-endpoint P99
latency for shift, reverse, and tornado
(Table~\ref{tab:route-overlap}). Tornado provides the clearest
example of the limitations of average hop count: the
high-overlap route set is slightly shorter on average
(44.0 versus 44.5 hops), yet worst-endpoint P99 latency rises
from 179 to 210 cycles, an increase of 17.3\%. Hotspot exhibits
the opposite ordering: average route length decreases from
37.0 to 36.0 hops, while worst-endpoint P99 decreases slightly
from 302 to 298 cycles. Thus, higher route overlap does not
uniformly imply higher tail latency across traffic patterns.
For shift and reverse, the higher-overlap routes are also
longer, so these comparisons do not separate the effects of
route length and resource sharing.

These results show that lower average hop count does not guarantee
lower worst-endpoint tail latency. They are
consistent with contention on shared NoC resources contributing
to latency, but do not isolate its contribution. Complementing
the AES-to-HBM placement study, this experiment demonstrates how
Naviq's configuration sweeps and per-endpoint latency measurements
support route selection with endpoint placement held fixed.

\subsection{Routing Strategy Comparison Study}
\label{sec:routing-congestion}

We compare four route-selection strategies under tornado traffic
to assess how hop count and routing affect packet tail latency.
Endpoint placement, source-to-destination mapping, and
traffic-generator settings are fixed. Routing strategies include
shortest-path, path-diverse that reduces global overlap while limiting additional hops, and deliberately
high-overlap and bad-path routes.

Each of four AXI4-MM traffic generators issues 5,000 reads and
 writes per seed using 32-byte beats and 16-beat bursts,
with configured rates of 800~MB/s per source in each direction.
All strategies use a 544-byte read burst allowance, a maximum
of 16 outstanding reads, and the default NoC-interface
buffer capacity of 32 entries.

For each source and direction, we pool 15,000 transactions
across three matched random seeds and compute P99 latency.
Table~\ref{tab:routing-strategies} reports the maximum
per-source P99 read and write latencies. The bandwidth reaches approximately 3.20~GB/s per direction
for all strategies, calculated by summing per-source rates
and averaging across seeds. Each source's rate is measured
from its first accepted request to its last completion.

\begin{table}[t]
  \centering
\caption{Average hop count and worst-endpoint P99 
latencies in cycles across routing strategies under tornado traffic.}
\label{tab:routing-strategies}
  \small
  \setlength{\tabcolsep}{5pt}
  \begin{tabular}{lrrr}
    \toprule
    Strategy & Avg. hops & Write P99 & Read P99 \\
    \midrule
    Shortest     & 43.0 & 209 & 199 \\
    Bad path     & 45.0 & 214 & 205 \\
    High overlap & 44.0 & 214 & 198 \\
    Path-diverse & 44.5 & 179 & 180 \\
    \bottomrule
  \end{tabular}
\end{table}

Relative to shortest-path routing, path-diverse routing
increases average route length from 43.0 to 44.5 hops while
reducing worst-endpoint write P99 from 209 to 179 cycles
(14.4\%) and read P99 from 199 to 180 cycles (9.5\%).
High-overlap routing increases write P99 while leaving
read P99 nearly unchanged, illustrating different effects
on read and write tail latency.

\begin{figure}[t]
  \centering
  \includegraphics[width=\columnwidth]
    {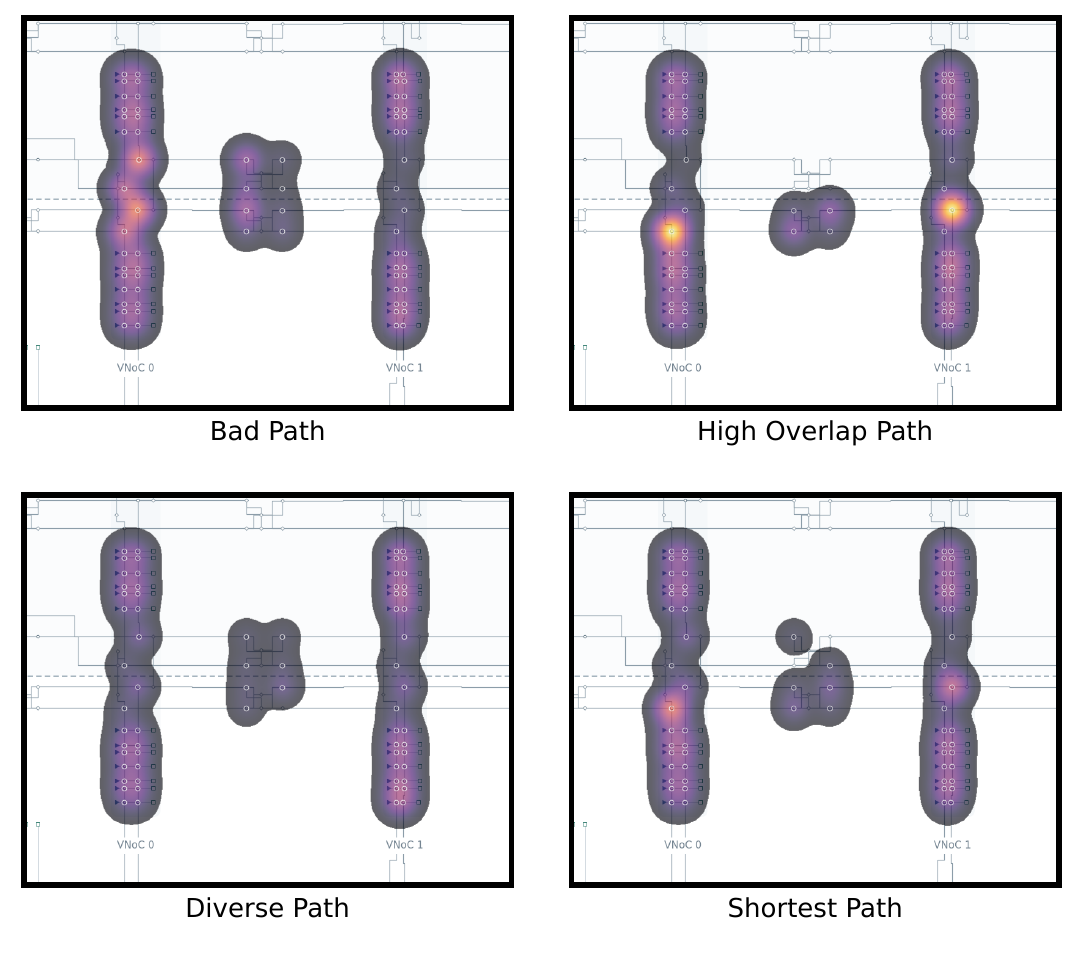}
  \caption{Buffer Occupancy Across Routing Strategies}
  \label{fig:routing-strategy-hotspots}
\end{figure}

Figure~\ref{fig:routing-strategy-hotspots} shows spatially
smoothed, time-averaged router buffer occupancy on a shared
color scale, highlighting where buffered traffic accumulates. These results demonstrate that lower average hop count does not
guarantee lower worst-endpoint tail latencies.

\subsection{NoC Deadlock Study}

\begin{figure}
    \centering
    \raisebox{-0.5\height}{%
        \includegraphics[width=0.73\columnwidth]
            {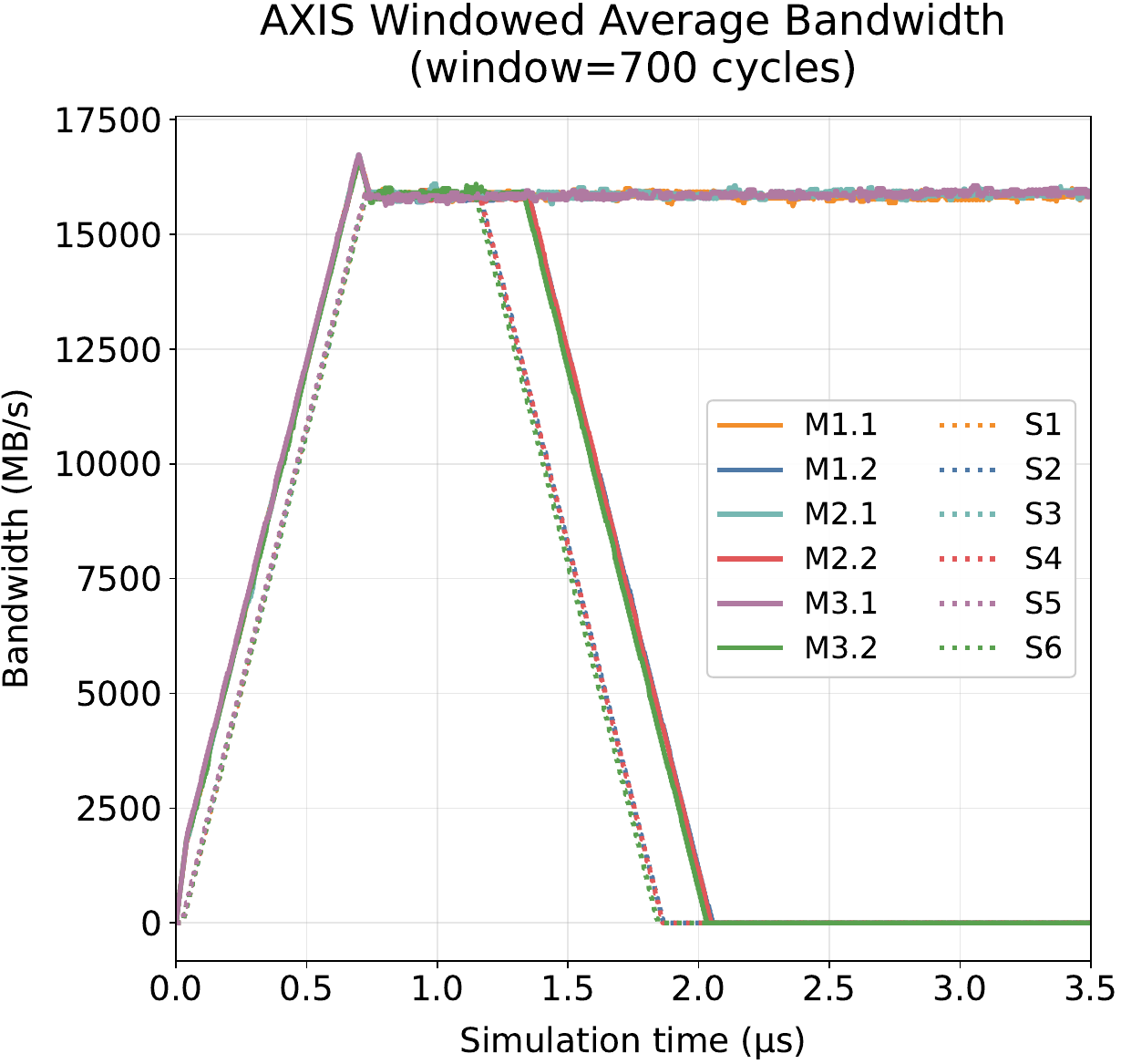}%
    }%
    \hspace{0.01\columnwidth}%
    \raisebox{-0.5\height}{%
        \includegraphics[width=0.08\columnwidth]
            {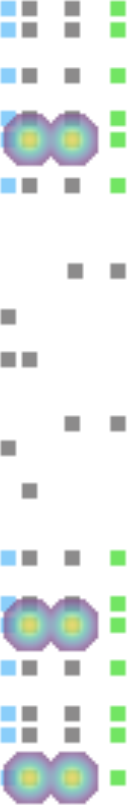}%
    }%
    \hspace{0.01\columnwidth}%
    \raisebox{-0.5\height}{%
        \includegraphics[width=0.12\columnwidth]
            {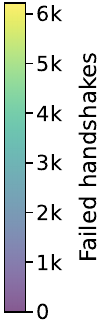}%
    }
    \caption{NoC Bandwidth and Failed Handshake Heatmap}
    \label{fig:debugging_plots}
\end{figure}

\begin{figure}
    \centering
    \includegraphics[width=0.85\columnwidth]{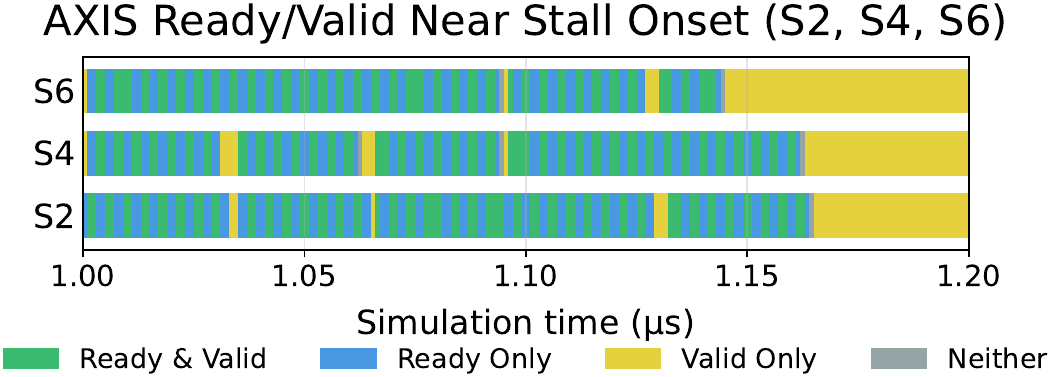}
    \caption{AXI4-Stream Ready/Valid Signals}
    \label{fig:debugging_stalls}
\end{figure}

Naviq provides integrated debugging support through protocol diagnostics, performance logging, activity visualization, and packet tracing. To demonstrate how these capabilities work together, we present a heterogeneous accelerator debugging case study. The experimental setup is highlighted in yellow in Figure~\ref{fig:experiment-setup}, with an enlarged view shown on the right. A control unit configures the accelerators through AXI4-MM connections labeled ‘C,’ whose paths are omitted for clarity, while three masters (M1–M3) stream data through AXI4-S to a downstream slave (S). Three of the six NSUs in S are synchronized using shared logic with receiver FIFOs that align packet arrivals. We intentionally inject a functional bug into one synchronization FIFO, but assume its exact location is unknown.

During simulation, Naviq first reports that multiple AXI4-S transactions have remained undelivered for over 5,000 cycles, indicating a potential deadlock. It then flags abnormal NPS activity by identifying switches with significantly higher or lower dequeue activity than the median. Guided by these warnings, the user examines Naviq's generated bandwidth plots, one of which is shown on the left of Figure~\ref{fig:debugging_plots}. In this plot, several links exhibit sharp bandwidth drops approximately 1.5 $\mu$s into the experiment.

Naviq also notifies the user of potential system-level congestion and provides several congestion-analysis plots. For example, Naviq reports average NPS queue occupancy and identifies how often and why AXI4 handshakes fail to complete. In this case, the failed handshake heatmap on the right of Figure \ref{fig:debugging_plots} is the most relevant to the issue. It reveals three persistent backpressure paths that result in a significant number of failed AXI4 handshakes. By cross-referencing these paths with the setup, the user identifies the affected links as M1.2, M2.2, and M3.2, all connected through the synchronization logic. Finally, Naviq's per-link ready/valid graph in Figure~\ref{fig:debugging_stalls} narrows down the problem further. It demonstrates that S6 lowers \textit{ready} before S4 and S2, which induces backpressure in the NoC and suggests that the fault originates near S6.

RTL inspection confirms faulty full/empty pointer logic in the S6 receiver FIFO. The FIFO incorrectly wraps its pointer state and permanently deasserts \texttt{TREADY}; backpressure propagates through the synchronized paths and produces a system-wide deadlock. Although the bug is confined to a small RTL module, its effects spread across the NoC and are difficult to localize from waveforms alone. Misbehaving accelerators can cause system-wide failures.

\subsection{HBM Performance Inefficiency Study}
Some bugs produce functionally correct but poorly performing hardware. To demonstrate Naviq's ability to diagnose such performance bugs, we examine a case study with an inefficient HBM access pattern that limits bandwidth.

The HBM access pattern is generated by a node that connects to the NoC through 32 HBM NMUs (highlighted in purple in Figure \ref{fig:experiment-setup}), each able to make its own AXI4-MM requests. Four HBM NMUs can connect to one HBM Memory Controller. The HBM Memory Controller divides four streams among two pseudochannels, each of which connects to two NoC slave ports that, in turn, each contain one HBM memory port. Each HBM bank can only service one row at a time; switching rows incurs latency. Achieved HBM bandwidth significantly benefits from row reuse \cite{amd_pg313_noc}.

\begin{figure}
    \centering
    \includegraphics[width=\columnwidth]{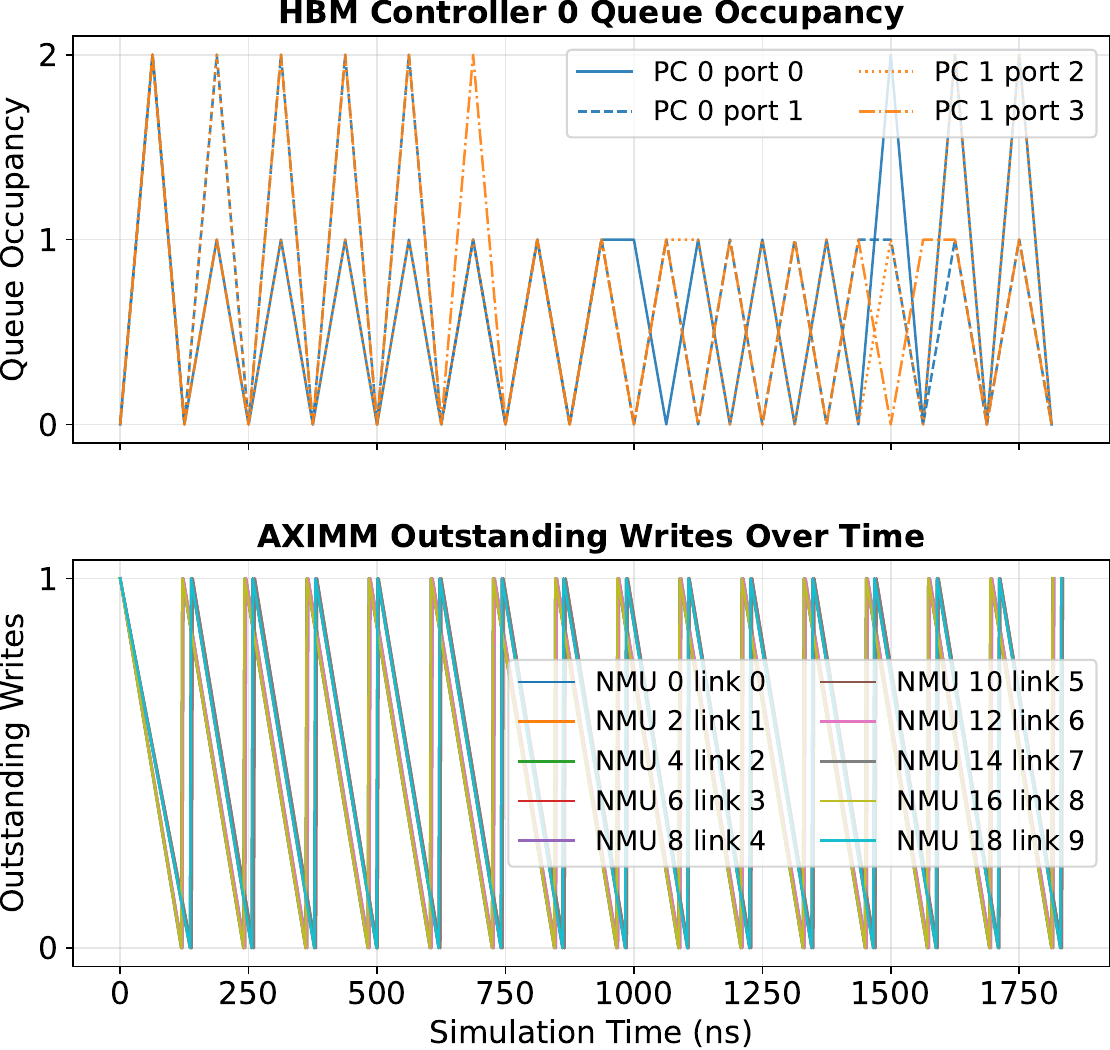}
    \caption{HBM Performance Inefficiencies Due to Low Queue Occupancies and Outstanding Write Counts}
    \label{fig:hbm_graphs}
\end{figure}

While simulating this buggy design, Naviq reports several potential indicators of inefficient memory accesses: low traffic utilization relative to the setup's theoretical maximum, high row-miss rates, and unbalanced traffic across HBM ports. When observed alongside low bandwidth, these warnings provide essential clues for diagnosing the performance bottleneck. Naviq's generated HBM controller queue occupancy (Figure \ref{fig:hbm_graphs} upper) and AXI4-MM outstanding write graphs (Figure \ref{fig:hbm_graphs} lower) reveal the issue. Very few operations are enqueued within the HBM controller and only a single outstanding write is permitted per NMU. These observations suggest that the bandwidth limitation may originate from the node rather than the NoC or memory subsystem.

Further investigation identifies the two contributing faults that we introduced into the configuration. First, the address-generation logic produces consecutive requests that access different rows in the same bank, reducing row locality. Second, each AXI4-MM generator is configured to support a single outstanding write. Together, these constraints reduce memory-level parallelism provided by HBM.

Without Naviq's integrated diagnostics, identifying the root cause of this performance bug would require manual investigation into data movement in multiple different components. Naviq's plots, diagnostic warnings, and other tooling narrow the search space quickly, helping users rapidly identify and localize problems.

\section{Conclusion}
\label{sec:conclusion}
Naviq is a fast, cycle-level, open-source software simulator for modern FPGA SoCs, with a focus on modeling real SoC platform behavior. Its hybrid C++/RTL approach delivers 81$\times$--182$\times$ speedup over vendor tools while preserving key NoC, memory, and accelerator interactions. Naviq's automated RTL integration flow enables rapid attachment and testing of large, multi-accelerator designs, while its visualization, sweeping, and analysis tools support full-system debugging and design-space exploration. Our repository can be found here: \url{https://github.com/UIUC-ChenLab/Naviq}.

\begin{acks}
This work was supported by the AMD Center of Excellence and AMD's  Heterogeneous Accelerated Compute Clusters (HACC) program. ChatGPT \cite{chatgpt} and Cursor \cite{cursor_ai} were used to help debug our simulator, write scripts, and assist with paper grammar and editing. 
\end{acks}

% \printbibliography
\bibliographystyle{ACM-Reference-Format}
\bibliography{citations}

@manual{AMBA,
title = {AMBA AXI and ACE Protocol Specification},
year = {2013},
author = {Arm},
url = {https://developer.arm.com/documentation/ihi0022/e},
pagetotal={328}
}

@manual{AMBA_AXIS,
title = {AMBA AXI-Stream Protocol Specification},
year = {2021},
author = {Arm},
url = {https://developer.arm.com/documentation/ihi0051/b/?lang=en},
pagetotal={56}
}

@manual{amd_pg313_noc,
  author       = {{Advanced Micro Devices, Inc.}},
  title        = {Versal Adaptive SoC Network on Chip (NoC)},
  year         = {2026},
  number       = {PG313},
  organization = {AMD},
  url          = {https://docs.amd.com/r/en-US/pg313-network-on-chip},
  note         = {Product guide for the Versal AXI Network-on-Chip IP}
}

@misc{amd_vivado,
  author       = {{Advanced Micro Devices, Inc.}},
  title        = {Vivado Design Suite},
  year         = {2026},
  howpublished = {\url{https://www.amd.com/en/products/software/adaptive-socs-and-fpgas/vivado.html}},
  note         = {FPGA design and verification tool},
}

@misc{verilator,
  author       = {Wilson Snyder},
  title        = {Verilator},
  year         = {2026},
  howpublished = {\url{https://www.veripool.org/verilator/}},
  note         = {Open-source cycle-accurate SystemVerilog simulator},
}

@techreport{asanovic2016rocket,
  title       = {The Rocket Chip Generator},
  author      = {Asanovi{\'c}, Krste and Avizienis, Rimas and Bachrach, Jonathan and Beamer, Scott and Biancolin, David and Celio, Christopher and Cook, Henry and Dabbelt, David and Hauser, John and Izraelevitz, Adam and Karandikar, Sagar and Keller, Ben and Kim, Donggyu and Koenig, John and Lee, Yunsup and Love, Jonathan and Maas, Martin and Magyar, Albert and Mao, Howard and Moreto, Miquel and Ou, Albert and Patterson, David A. and Richards, Brian and Schmidt, Colin and Twigg, Stephen and Vo, Huy and Waterman, Andrew},
  institution = {EECS Department, University of California, Berkeley},
  number      = {UCB/EECS-2016-17},
  year        = {2016},
  month       = apr,
  url         = {https://www2.eecs.berkeley.edu/Pubs/TechRpts/2016/EECS-2016-17.pdf}
}

@misc{opentitan,
  title        = {OpenTitan: An Open Source, High-Quality Silicon Root of Trust},
  author       = {{lowRISC CIC}},
  howpublished = {\url{https://opentitan.org}},
  year         = {2026}
  }

@article{gem5,
author = {Binkert, Nathan and Beckmann, Bradford and Black, Gabriel and Reinhardt, Steven K. and Saidi, Ali and Basu, Arkaprava and Hestness, Joel and Hower, Derek R. and Krishna, Tushar and Sardashti, Somayeh and Sen, Rathijit and Sewell, Korey and Shoaib, Muhammad and Vaish, Nilay and Hill, Mark D. and Wood, David A.},
title = {The gem5 simulator},
year = {2011},
issue_date = {May 2011},
publisher = {Association for Computing Machinery},
address = {New York, NY, USA},
volume = {39},
number = {2},
issn = {0163-5964},
url = {https://doi.org/10.1145/2024716.2024718},
doi = {10.1145/2024716.2024718},
journal = {SIGARCH Comput. Archit. News},
month = aug,
pages = {1–7},
numpages = {7}
}

@INPROCEEDINGS{garnet,
  author={Agarwal, Niket and Krishna, Tushar and Peh, Li-Shiuan and Jha, Niraj K.},
  booktitle={2009 IEEE International Symposium on Performance Analysis of Systems and Software}, 
  title={GARNET: A detailed on-chip network model inside a full-system simulator}, 
  year={2009},
  volume={},
  number={},
  pages={33-42},
  doi={10.1109/ISPASS.2009.4919636}}

@misc{amd_alveo_v80,
  author       = {{Advanced Micro Devices, Inc.}},
  title        = {{Alveo V80 Data Center Accelerator Card}},
  year         = {2026},
  howpublished = {\url{https://www.amd.com/en/products/accelerators/alveo/v80.html}}
}

@misc{amd_pg453_axi_switch,
  author       = {{Advanced Micro Devices, Inc.}},
  title        = {{AXI Switch v1.0 LogiCORE IP Product Guide (PG453)}},
  year         = {2026},
  howpublished = {\url{https://docs.amd.com/r/en-US/pg453-axi-switch}}
}

@INPROCEEDINGS{7372595,
  author={Cong, Jason and Fang, Zhenman and Gill, Michael and Reinman, Glenn},
  booktitle={2015 IEEE/ACM International Conference on Computer-Aided Design (ICCAD)}, 
  title={PARADE: A cycle-accurate full-system simulation Platform for Accelerator-Rich Architectural Design and Exploration}, 
  year={2015},
  volume={},
  number={},
  pages={380-387},
  doi={10.1109/ICCAD.2015.7372595}}

@INPROCEEDINGS{6853196,
  author={Shao, Yakun Sophia and Reagen, Brandon and Wei, Gu-Yeon and Brooks, David},
  booktitle={2014 ACM/IEEE 41st International Symposium on Computer Architecture (ISCA)}, 
  title={Aladdin: A pre-RTL, power-performance accelerator simulator enabling large design space exploration of customized architectures}, 
  year={2014},
  volume={},
  number={},
  pages={97-108},
  doi={10.1109/ISCA.2014.6853196}}

@INPROCEEDINGS{7783751,
  author={Shao, Yakun Sophia and Xi, Sam Likun and Srinivasan, Vijayalakshmi and Wei, Gu-Yeon and Brooks, David},
  booktitle={2016 49th Annual IEEE/ACM International Symposium on Microarchitecture (MICRO)}, 
  title={Co-designing accelerators and SoC interfaces using gem5-Aladdin}, 
  year={2016},
  volume={},
  number={},
  pages={1-12},
  doi={10.1109/MICRO.2016.7783751}}

@inproceedings{10.1145/3472456.3472461,author = {L\'{o}pez-Parad\'{\i}s, Guillem and Armejach, Adri\`{a} and Moret\'{o}, Miquel},title = {gem5 + rtl: A Framework to Enable RTL Models Inside a Full-System Simulator},year = {2021},isbn = {9781450390682},publisher = {Association for Computing Machinery},address = {New York, NY, USA},url = {https://doi.org/10.1145/3472456.3472461},doi = {10.1145/3472456.3472461},booktitle = {Proceedings of the 50th International Conference on Parallel Processing},articleno = {29},numpages = {11},location = {Lemont, IL, USA},series = {ICPP '21}}

@INPROCEEDINGS{heterogarnet,
  author={Bharadwaj, Srikant and Yin, Jieming and Beckmann, Bradford and Krishna, Tushar},
  booktitle={2020 57th ACM/IEEE Design Automation Conference (DAC)}, 
  title={Kite: A Family of Heterogeneous Interposer Topologies Enabled via Accurate Interconnect Modeling}, 
  year={2020},
  volume={},
  number={},
  pages={1-6},
  doi={10.1109/DAC18072.2020.9218539}}

@misc{lowepower2020gem5simulatorversion200,
      title={The gem5 Simulator: Version 20.0+}, 
      author={Jason Lowe-Power and Abdul Mutaal Ahmad and Ayaz Akram and Mohammad Alian and Rico Amslinger and Matteo Andreozzi and Adrià Armejach and Nils Asmussen and Brad Beckmann and Srikant Bharadwaj and Gabe Black and Gedare Bloom and Bobby R. Bruce and Daniel Rodrigues Carvalho and Jeronimo Castrillon and Lizhong Chen and Nicolas Derumigny and Stephan Diestelhorst and Wendy Elsasser and Carlos Escuin and Marjan Fariborz and Amin Farmahini-Farahani and Pouya Fotouhi and Ryan Gambord and Jayneel Gandhi and Dibakar Gope and Thomas Grass and Anthony Gutierrez and Bagus Hanindhito and Andreas Hansson and Swapnil Haria and Austin Harris and Timothy Hayes and Adrian Herrera and Matthew Horsnell and Syed Ali Raza Jafri and Radhika Jagtap and Hanhwi Jang and Reiley Jeyapaul and Timothy M. Jones and Matthias Jung and Subash Kannoth and Hamidreza Khaleghzadeh and Yuetsu Kodama and Tushar Krishna and Tommaso Marinelli and Christian Menard and Andrea Mondelli and Miquel Moreto and Tiago Mück and Omar Naji and Krishnendra Nathella and Hoa Nguyen and Nikos Nikoleris and Lena E. Olson and Marc Orr and Binh Pham and Pablo Prieto and Trivikram Reddy and Alec Roelke and Mahyar Samani and Andreas Sandberg and Javier Setoain and Boris Shingarov and Matthew D. Sinclair and Tuan Ta and Rahul Thakur and Giacomo Travaglini and Michael Upton and Nilay Vaish and Ilias Vougioukas and William Wang and Zhengrong Wang and Norbert Wehn and Christian Weis and David A. Wood and Hongil Yoon and Éder F. Zulian},
      year={2020},
      eprint={2007.03152},
      archivePrefix={arXiv},
      primaryClass={cs.AR},
      url={https://arxiv.org/abs/2007.03152}, 
}

@INPROCEEDINGS{PAAS,
  author={Liang, Tingyuan and Feng, Liang and Sinha, Sharad and Zhang, Wei},
  booktitle={2017 27th International Conference on Field Programmable Logic and Applications (FPL)}, 
  title={PAAS: A system level simulator for heterogeneous computing architectures}, 
  year={2017},
  volume={},
  number={},
  pages={1-8},
  doi={10.23919/FPL.2017.8056775}}

@ARTICLE{gem5-gpu,
  author={Power, Jason and Hestness, Joel and Orr, Marc S. and Hill, Mark D. and Wood, David A.},
  journal={IEEE Computer Architecture Letters}, 
  title={gem5-gpu: A Heterogeneous CPU-GPU Simulator}, 
  year={2015},
  volume={14},
  number={1},
  pages={34-36},
  doi={10.1109/LCA.2014.2299539}}

@INPROCEEDINGS{gpgpusim_2,
  author={Bakhoda, Ali and Yuan, George L. and Fung, Wilson W. L. and Wong, Henry and Aamodt, Tor M.},
  booktitle={2009 IEEE International Symposium on Performance Analysis of Systems and Software}, 
  title={Analyzing CUDA workloads using a detailed GPU simulator}, 
  year={2009},
  volume={},
  number={},
  pages={163-174},
  doi={10.1109/ISPASS.2009.4919648}}

@INPROCEEDINGS{gpgpusim_1,
  author={Fung, Wilson W.L. and Sham, Ivan and Yuan, George and Aamodt, Tor M.},
  booktitle={40th Annual IEEE/ACM International Symposium on Microarchitecture (MICRO 2007)}, 
  title={Dynamic Warp Formation and Scheduling for Efficient GPU Control Flow}, 
  year={2007},
  volume={},
  number={},
  pages={407-420},
  doi={10.1109/MICRO.2007.30}}

@misc{cursor_ai,
  author       = {{Anysphere}},
  title        = {Cursor: AI-Powered Code Editor},
  howpublished = {\url{https://www.cursor.com}},
  year         = {2026}
}

@misc{chatgpt,
  author       = {{OpenAI}},
  title        = {{ChatGPT}},
  howpublished = {\url{https://chat.openai.com}},
  year = {2026}
}

@INPROCEEDINGS {RADSim,
author = { Boutros, Andrew and Nurvitadhi, Eriko and Betz, Vaughn },
booktitle = { 2022 32nd International Conference on Field-Programmable Logic and Applications (FPL) },
title = {{RAD-Sim: Rapid Architecture Exploration for Novel Reconfigurable Acceleration Devices}},
year = {2022},
volume = {},
ISSN = {},
pages = {438-444},
doi = {10.1109/FPL57034.2022.00072},
url = {https://doi.ieeecomputersociety.org/10.1109/FPL57034.2022.00072},
publisher = {IEEE Computer Society},
address = {Los Alamitos, CA, USA},
month =sep}

@misc{liu2025demystifyingfpgahardnoc,
      title={Demystifying FPGA Hard NoC Performance}, 
      author={Sihao Liu and Jake Ke and Tony Nowatzki and Jason Cong},
      year={2025},
      eprint={2503.10861},
      archivePrefix={arXiv},
      primaryClass={cs.AR},
      url={https://arxiv.org/abs/2503.10861}, 
}

@INPROCEEDINGS {booksim,
author = { Jiang, Nan and Balfour, James and Becker, Daniel U. and Towles, Brian and Dally, William J. and Michelogiannakis, George and Kim, John },
booktitle = { IEEE International Symposium on Performance Analysis of Systems and Software (ISPASS 2013) },
title = {{A detailed and flexible cycle-accurate Network-on-Chip simulator}},
year = {2013},
volume = {},
ISSN = {},
pages = {86-96},
doi = {10.1109/ISPASS.2013.6557149},
url = {https://doi.ieeecomputersociety.org/10.1109/ISPASS.2013.6557149},
publisher = {IEEE Computer Society},
address = {Los Alamitos, CA, USA},
month =apr}

@misc{amd_smartconnect,
  author       = {{Advanced Micro Devices, Inc.}},
  title        = {{AXI SmartConnect}},
  howpublished = {\url{https://docs.amd.com/r/en-US/pg247-smartconnect/}},
  year = {2026}
}

@inproceedings{PANIC,
author = {Lin, Jiaxin and Patel, Kiran and Stephens, Brent E. and Sivaraman, Anirudh and Akella, Aditya},
title = {PANIC: a high-performance programmable NIC for multi-tenant networks},
year = {2020},
isbn = {978-1-939133-19-9},
publisher = {USENIX Association},
address = {USA},
booktitle = {Proceedings of the 14th USENIX Conference on Operating Systems Design and Implementation},
articleno = {14},
numpages = {17},
series = {OSDI'20}
}

@inproceedings{SuperNIC,
author = {Lin, Will and Shan, Yizhou and Kosta, Ryan and Krishnamurthy, Arvind and Zhang, Yiying},
title = {SuperNIC: An FPGA-Based, Cloud-Oriented SmartNIC},
year = {2024},
isbn = {9798400704185},
publisher = {Association for Computing Machinery},
address = {New York, NY, USA},
url = {https://doi.org/10.1145/3626202.3637564},
doi = {10.1145/3626202.3637564},
booktitle = {Proceedings of the 2024 ACM/SIGDA International Symposium on Field Programmable Gate Arrays},
pages = {130–141},
numpages = {12},
location = {Monterey, CA, USA},
series = {FPGA '24}
}

@techreport{nist2001aes,
  author      = {{National Institute of Standards and Technology}},
  title       = {{Advanced Encryption Standard (AES)}},
  institution = {National Institute of Standards and Technology},
  type        = {Federal Information Processing Standards Publication},
  number      = {197},
  year        = {2001},
  month       = nov,
  url         = {https://csrc.nist.gov/files/pubs/fips/197/final/docs/fips-197.pdf}
}

@misc{secworks_aes,
  author       = {{secworks}},
  title        = {{AES}: Verilog Implementation of the Advanced Encryption Standard},
  howpublished = {\url{https://github.com/secworks/aes}},
  note         = {Verilog implementation of AES as specified in NIST FIPS 197},
  year         = {2026}
}
%%
%% If your work has an appendix, this is the place to put it.
% \appendix
\end{document}